\documentclass[
aip, 
apl, 
final, 
preprint, 
reprint, 
showpacs, 
amsmath, 
amssymb]{revtex4-1}
\usepackage{graphicx}
\usepackage{dcolumn}
\usepackage{bm}
\usepackage{textcomp}
\usepackage{revsymb}
\usepackage{amssymb}
\usepackage{ifsym}
\usepackage{wasysym}
\usepackage{stmaryrd}
\usepackage{txfonts}

\begin{document}


\title{Room-temperature formation of Pt$_3$Si/Pt$_2$Si films on poly-Si substrates}
 
\author{V. P. Dubkov}
\email{vladub@kapella.gpi.ru}
\noaffiliation

\author{S. A. Mironov}
\noaffiliation

\author{K. V. Chizh} 
\noaffiliation

\author{V. A. Yuryev}
\email{vyuryev@kapella.gpi.ru} 
\noaffiliation

\affiliation{A.\,M.\,Prokhorov General Physics Institute of the Russian Academy of Sciences, 38 Vavilov Street, Moscow, 119991, Russia}

\date{\today}%

\begin{abstract}
We propose a way of formation of thin bilayer Pt$_3$Si/Pt$_2$Si films at room temperature on poly-Si substrates by Pt magnetron sputtering and wet etching, obtain such film, investigate its structure and phase composition and estimate the thickness of its layers.
We verify by direct x-ray photoelectron-spectroscopic measurements our previous observation of the Pt$_2$Si layer formaton between Pt and poly-Si films as a result of Pt magnetron sputtering at room temperature. This layer likely appears due to high enough temperature of Pt ions in the magnetron plasma sufficient for chemical reaction of the silicide film formation on the Si surface. 
The Pt$_3$Si layer likely forms from the Pt--Pt$_3$Si layer (Pt$_{95}$Si$_5$), which arises under Pt film during the magnetron sputtering, as a result of Pt removal by wet etching.
\end{abstract}



\maketitle

Platinum silicides have attracted attention of researchers for a number of decades due to their exceptional prospectiveness in microelectronics and silicon-based microphotonics.\cite{Silicides,Murarka,*Murarka-1995}
PtSi infrared detector arrays represented a qualitative breakthrough and  opened a new era in the infrared imaging technology.\cite{Rogalski_3,*Kimata_2-PtSi}
Now Pt silicides are considered as metals for Schottky-barrier formation to poly-Si:P in thin-film diode  bolometers.\cite{PtSi_Schottky-diodes_bolometers, SPIE_PtSi-bolometer-cite} 
There is no doubt about the prospectiveness of their application as ohmic contacts and submicron lines in microelectronics especially taking into account their low formation temperatures that is very important for CMOS and especially for nanoelectronic devices.\cite{PtSi_in_submicron_lines,*Silicides_ohmic_contacts,*VCIAN2011}
In addition, low-temperature silicides, e.g. Pt$_2$Si, are expected to be  suitable for formation of uniform Schottky contacts in devices of power electronics in which the sizes of contacts reach millimeters and nonuniformity of the barrier height causes a significant increase in reverse current.\cite{Pt2Si-PtSi_Belarus}
Pt silicide ohmic contacts as well as Schottky barrier formation on polycrystalline\cite{Silicides_Polysilicon} or amorphous silicon is of special interest for photovoltaic and sensor technology.\cite{SPIE_PtSi-bolometer-cite}
So, this class of materials should be considered as one of the most friendly ones to silicon technology.  
However, as the used silicide films become thinner and reach tens or even units of nanometers their resistivity becomes of primary importance. In this connection  Pt$_3$Si, which has the lowest  sheet resistance among the Pt silicides (18.9 $\Omega/\boxempty$ in comparison with 2.6 $\Omega/\boxempty$ of Pt, 31.8 $\Omega/\boxempty$ of Pt$_2$Si  and 57.6 $\Omega/\boxempty$ of PtSi),\cite{Streller_Pt3Si-2016} starts to play the main role.
So, the development of simple processes of formation of thin Pt$_3$Si films becomes more and more important.

This letter presents a simple CMOS compatible process of a thin bilayer Pt$_3$Si/Pt$_2$Si film formation at room temperature on a polycrystalline silicon substrate.

\begin{figure}[b]
\includegraphics[scale=.56]{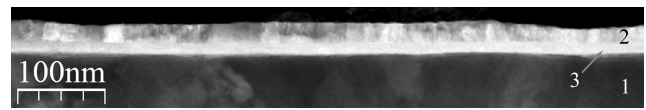}
\caption{\label{fig:STEM}
STEM image of the as-deposited Pt/poly-Si structure shows a poly-Si layer (1), a layer of Pt  (2) and an interfacial layer (3) consisting of a Pt$_{95}$Si$_5$  layer 
on a layer of Pt$_2$Si.\cite{PtSi_Schottky-diodes_bolometers}
}
\end{figure}

A 35-nm 
thick film of platinum was deposited by magnetron sputtering at room temperature on a 125-nm thick poly-Si:P layer formed on a Si$_3$N$_4$/SiO$_2$/Si(001) artificial substrate (Fig.\,\ref{fig:STEM}). After deposition, platinum was removed by chemical etching in a warm aqueous solution of \textit{aqua regia}  (H$_2$O\,:\,HCl\,:\,HNO$_3$ [4\,:\,3\,:\,1]).\footnote{
The 
reaction is
3Pt + 18HCl + 4HNO$_3 \rightarrow $ 3H$_2$[PtCl$_6$] + 4NO${\uparrow}$ + 8H$_2$O.
}
Details of the sample preparation process can be found in Ref.\,\onlinecite{PtSi_Schottky-diodes_bolometers}.

\begin{figure}[t]
\includegraphics[scale=.55]{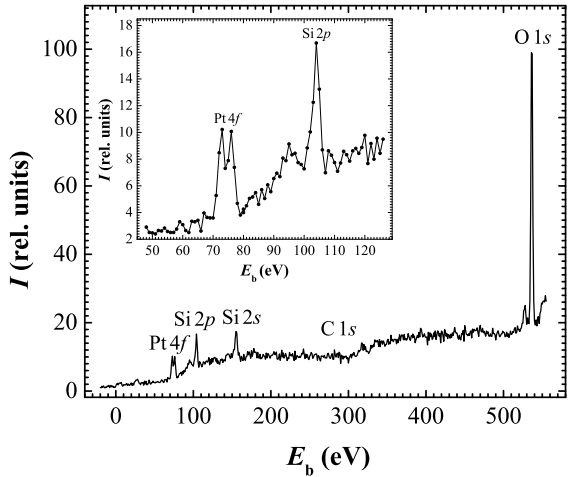}
\caption{\label{fig:XPS}
Survey XPS spectrum of the  Pt-silicide/poly-Si film. The insert demonstrates a magnified region around the Pt\,4$f$ and Si\,2$p$ peaks. 
}
\end{figure}

The samples  were studied by means of the  X-ray photoelectron spectroscopy (XPS).
The measurements were carried out using a cylindrical mirror electron energy analyser\cite{Cylindrical_Mirror_Analyzer} (Riber\,EA\,150) installed in the ultrahigh-vacuum
chamber; 
the residual gas pressure in the chamber did not exceed $7\times 10^{-7}$\,Pa.  
Non-monochromatic
Al\,K$_{\alpha}$ x-rays ($\hbar\omega = 1486.7$\,eV) were used for photoexcitation of electrons. 
Survey spectra were scanned at the resolution (FWHM) better than 1.8\,eV; 
high resolution spectra of specific elements were obtained at the resolution not worse than 0.96\,eV.
XPSPEAK\,4.1 peak fitting program was utilized for treatment of spectra.
It was taken into the account during peak deconvolution that the Al\,K$_{\alpha}$ band consists of two lines, K$_{\alpha_1}$ and  K$_{\alpha_2}$, with energy difference of $\hbar\Delta\omega = 0.5$\,eV.
Shifts of peaks related to elements in chemical compounds were compared with the NIST X-ray Photoelectron Spectroscopy (XPS) Database.\cite{XPS_NIST_Database, NIST_XPS_Database}
Relative concentrations of atoms were estimated from ratios of normalized areas under corresponding peaks.
Inelastic mean free path (IMFP) was estimated using TPP-2M equation and the NIST Database.\cite{NIST-IMFP, NIST_IMFP_Database}
The spectrometer was calibrated against the Si$^{4+}$ signal ($E_{\mathrm b}=103.6$ eV)  since the C\,1$s$ XPS peak recorded at high resolution had low signal-to-noise ratio (Fig.\,\ref{fig:XPS}) that did not allow us to obtain reliable data on its energy position. 

The scanning transmission electron microscopy (STEM) image was obtained using the Carl Zeiss Libra-200 FE HR transmission electron microscope; the WSxM software was used for image processing.\cite{WSxM}


Fig.\,\ref{fig:STEM} demonstrates a STEM image of the as-deposited Pt/poly-Si structure. This structures was previously analysed in details using STEM, x-ray diffraction and reflection, and XPS. It is composed by a poly-Si layer, a layer of Pt and an interfacial layer consisting of a Pt$_{95}$Si$_5$ layer formed 
on a layer of Pt$_2$Si.\cite{PtSi_Schottky-diodes_bolometers} This structure was subjected to wet etching to remove Pt.

Photoelectron peaks of Pt, Si, O and C are observed in the survey photoelectron spectrum of the  film obtained after Pt removal (Fig.\,\ref{fig:XPS}). Monotonic growth of signal is observed in the spectrum on the right (at the higher $E_{\mathrm b}$ side) of the Pt\,4$f$ doublet that is explained by inelastic scattering of the photoexcited  4$f$ electrons of Pt. This allows us to make a conclusion that the film containing Pt is overlaid by a layer of another composition in which the inelastic scattering happens.\cite{Tougaard}

\begin{figure}[t]
\includegraphics[scale=.55]{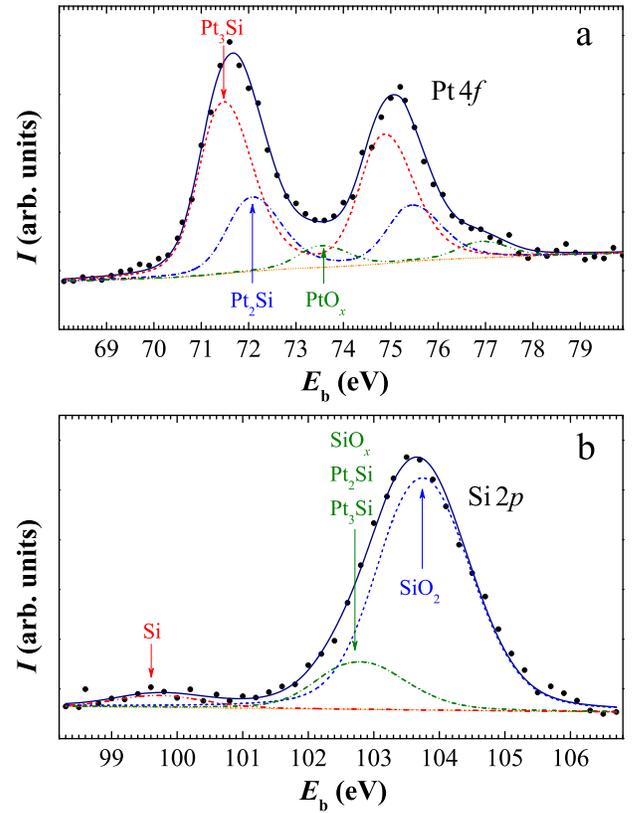}
\caption{\label{fig:XPS-HR}(Color online) 
High-resolution XPS spectra of Pt\,4$f$ (a) and Si\,2$p$ (b) obtained from the resultant Pt silicide film on the poly-Si substrate; measured bands are shown by solid lines, peaks obtained as a result of deconvolution are drawn by dotted and dashed lines
}
\end{figure}

A high-resolution photoelectron spectrum of Pt\,4$f$ consists of three components (Fig.\,\ref{fig:XPS-HR}a). The first one is described by a doublet with $E_{\mathrm b} = 71.5$\, eV and distance between peaks $\Delta E = 3.38$ eV which may be attributed to Pt$_3$Si;
the next one with $E_{\mathrm b} = 72.1$ eV and $\Delta E = 3.38$ eV obviously corresponds to Pt$_2$Si.\cite{NIST_XPS_Database, Streller_Pt3Si-2014,Streller_Pt3Si-2015,Fryer_Pt3Si-2016}
The third component lays at higher energies ($E_{\mathrm b} = 73.62$ eV, $\Delta E = 3.38$ eV) and may be attributed to both 
platinum oxide PtO$_x$ ($E_{\mathrm b} = 72.8$ to 74.6 eV)\footnote{\label{footnote:PtO} 
Likely PtO, $E_{\mathrm b} = 73.8$ eV, Ref.\,\onlinecite{NIST_XPS_Database}.
}
and 
PtCl$_2$ ($E_{\mathrm b} = 73.6$ eV)\cite{NIST_XPS_Database} 
or 
[Pt(NH$_3$)$_2$(NO$_2$)$_2$]\footnote{
CAS Registry No.\,14286023, Ref.\,\onlinecite{NIST_XPS_Database}.
}
 ($E_{\mathrm b} = 73.7$ eV)\cite{NIST_XPS_Database} which  may arise as a result of Pt etching in the \textit{aqua regia} solution. The ratio of peak areas for Pt$_3$Si, Pt$_2$Si and PtO$_x$ 
is  $64:28:8$.

The main contribution to the signal of silicon (Fig.\,\ref{fig:XPS-HR}b) is made by Si$^{4+}$ ($E_{\mathrm b} = 103.6$ eV) obviously related to silicon dioxide. In addition, a Si$^{3+}$ peak can be detected by deconvolution which can be explained by superposition of photoelectron peaks related to Pt$_3$Si, Pt$_2$Si ($E_{\mathrm b} = 102.6$ eV) and SiO$_x$ compounds having a broad range of binding energy values ($E_{\mathrm b} = 100.4$ to 103.6 eV)\cite{NIST_XPS_Database}.
An estimate of contribution  of Si atoms contained in Pt$_3$Si and Pt$_2$Si to this peak made in assumption of uniformity of the upper film and  taking into account empirical sensitivity factors of platinum and silicon gives the value of $\sim 16$\,\%.

Notice that an additional cycle of etching in the \textit{aqua regia} solution changed neither  a ratio of the photoelectron peak intensities nor their energy positions.

It should be noted also that a small photoelectron peak at $E_{\mathrm b} = 99.5$ eV is also observed in the Si\,2$p$ spectrum; it corresponds to pure Si of the polycrystalline substrate that allows us to roughly estimate the thickness of the formed film by 3 to 5 values of inelastic mean free path of electrons, i.e. by the value from 5 to 10 nm.

More accurate estimation of the film thickness can be made from the ratio of areas under the Pt\,4$f$ peaks of Pt$_3$Si and Pt$_2$Si $\sigmaup\approx 2.3$. 
Let us suppose that the Pt$_3$Si layer overlays the Pt$_2$Si one. (This assumption is based on the x-ray phase analysis according to which the Pt$_2$Si  and Pt$_{95}$Si$_5$ layers, 5.6 and 10\,nm thick, respectively, arise between poly-Si and Pt as a result of Pt sputtering.)\cite{PtSi_Schottky-diodes_bolometers} 
Considering the IMFP value\cite{NIST_IMFP_Database} of Pt$_3$Si we obtain the thickness of the Pt$_3$Si layer $h_1$ from the values of  densities and molar weights  of Pt$_3$Si and Pt$_2$Si, $\sigmaup$ and the thickness of the Pt$_2$Si layer $h_2$:\footnote{
The value of $h_1$ is calculated from a recursive equation:
$h_1 = h_{\mathrm{IMFP}}\,\times 
\ln[2 \sigmaup \muup_{\mathrm{Pt_3Si}}\varrhoup_{\mathrm{Pt_2Si}}h_2/3  \muup_{\mathrm{Pt_2Si}}\varrhoup_{\mathrm{Pt_3Si}}h_1]$,
where 
$h_{\mathrm{IMFP}}$
is the inelastic mean free path of electrons in Pt$_3$Si,\cite{NIST_IMFP_Database}
$\muup_{\mathrm{Pt_2Si}}$,
$\varrhoup_{\mathrm{Pt_2Si}}$,
$\muup_{\mathrm{Pt_3Si}}$
and
$\varrhoup_{\mathrm{Pt_3Si}}$
are molar weights and  densities  of Pt$_2$Si and Pt$_3$Si, respectively.
} 
$h_1 \approx 2.6$ nm. So, the total thickness of the bilayer Pt-silicide film $h_1+h_2\approx 8.2$ nm that is in good agreement with the above rough estimate.

Now, if we assume that the Pt$_{95}$Si$_5$ layer previously detected\cite{PtSi_Schottky-diodes_bolometers} under Pt is a mixture of Pt and  Pt$_3$Si (16Pt + Pt$_3$Si) we may suppose also that Pt is etched away from this layer during processing in the \textit{aqua regia} solution and the  Pt$_3$Si layer   is deposited atop Pt$_2$Si. We can estimate  $h_1$ from the thickness of Pt$_{95}$Si$_5$ and the densities of Pt, Pt$_{95}$Si$_5$ (20.5 g/cm$^3$)\cite{PtSi_Schottky-diodes_bolometers} and Pt$_3$Si:  $h_1 \approx 2.4$ nm. Thus, the silicide film thickness estimate of 8.0 nm obtained from the x-ray phase analysis\cite{PtSi_Schottky-diodes_bolometers} corresponds with that obtained from XPS.


In summary, we can make the following conclusions. 
(i)~We have verified our previous observation of Pt$_2$Si formation in between Pt and poly-Si films as a result of Pt magnetron sputtering at room temperature.\cite{PtSi_Schottky-diodes_bolometers} This phenomenon probably occurs due to high enough kinetic energy of Pt ions reaching the Si surface sufficient to give rise to chemical reaction of the silicide film formation. It opens a pathway to room-temperature process of formation of thin Pt silicide  films on Si.
(ii)~We have proposed a way of formation of thin bilayer Pt$_3$Si/Pt$_2$Si films on poly-Si substrates at room temperature by magnetron sputtering of Pt followed by etching in a warm aqueous solution of \textit{aqua regia} and obtained such films using this completely CMOS compatible process.
(iii)~Using the x-ray photoelectron spectroscopy we have verified the structure and composition of the formed Pt$_3$Si/Pt$_2$Si films; we have estimated the thickness of the Pt$_3$Si layer of the films and found it to be equal to $\sim 2.6$ nm; the thickness of the Pt$_3$Si/Pt$_2$Si films  has been found to be $\sim 8.2$ nm. The obtained estimates practically coincide with those made on the basis of our previous data of the  x-ray phase analysis.\cite{PtSi_Schottky-diodes_bolometers}
(iv)~We assume that the Pt$_3$Si layer is formed from the Pt$_{95}$Si$_5$ layer, which appears under Pt film during the magnetron sputtering,\cite{PtSi_Schottky-diodes_bolometers} as a result of Pt removal by wet etching.


This research was funded by RFBR (grant No.~16-32-00854).
Equipment of the Center for Collective Use of Scientific Equipment of GPI RAS was used for the study. 
We thank Ms. N.~V.~Kiryanova for her contribution to management of this research, Mr. O.~V.~Uvarov for obtaining the STEM images and Ms. L.~A.~Krylova for chemical treatments of the  samples.






%

\end{document}